\documentclass[galaxies,review,accept,pdftex,oneauthor]{Definitions/mdpi} 
\usepackage{graphicx}	
\usepackage{amsmath}	
\usepackage{amssymb}	
\usepackage{siunitx}
\firstpage{1} 
\pubvolume{1}
\issuenum{1}
\articlenumber{0}
\pubyear{2026}
\copyrightyear{2026}
\datereceived{ } 
\daterevised{ } 
\dateaccepted{ } 
\datepublished{ } 
\Title{Gradients of Turbulent Fluctuations as New Tracers of Astrophysical Magnetic Fields}

\Author{Yue Hu $^{1}$\orcidA{}*}

\AuthorNames{Yue Hu}

\address{%
$^{1}$ \quad Cahill Center for Astronomy and Astrophysics, California Institute of Technology, Pasadena, CA 91125, USA
}

\corres{Correspondence: yuehu@caltech.edu}

\abstract{Magnetohydrodynamic (MHD) turbulence is ubiquitous in astrophysical plasmas and produces scale-dependent anisotropy with respect to the local magnetic field. Because turbulence cascades preferentially perpendicular to the field, its spatial gradients acquire preferred orientations that can be used to probe magnetic-field structure. This property underlies a family of gradient techniques based on spectroscopic, synchrotron-intensity, and synchrotron-polarization observations. In this review, we summarize the theoretical foundation of these techniques, including the anisotropy of MHD turbulence and the corresponding behavior of velocity, magnetic-field, and density fluctuations. We then describe how these fluctuations are encoded in spectroscopic data and synchrotron observables, and review the implementation of Velocity Centroid Gradients, Velocity Channel Gradients, Intensity Gradients, Synchrotron Intensity Gradients, and Synchrotron Polarization Gradients. We also discuss the effects of isotropic fast modes, shocks, self-gravity, stellar outflows, thermal broadening, and self-absorption. Finally, we review applications to the diffuse interstellar medium, star-forming regions, the Galactic center, external galaxies, and galaxy clusters. 
}


\begin{document}

\section{Introduction}
\label{sec:introduction}

Magnetic fields are important across a wide range of astrophysical environments, from the diffuse interstellar medium (ISM) and star-forming molecular clouds to galactic nuclei, external galaxies, and the intracluster medium (ICM) \citep{Ferriere2001,Crutcher2012,Beck2015,CarilliTaylor2002,Han2017}. They regulate the transport of matter and energy, influence cosmic-ray propagation, mediate the interaction between turbulence and gravity, and contribute to processes ranging from star formation to galaxy evolution \citep{Ferriere2001,Crutcher2012,Beck2015}. Yet reconstructing astrophysical magnetic fields remains challenging. Polarized dust, starlight, and synchrotron emission constrain the plane-of-sky (POS) magnetic-field orientation, Faraday rotation probes the electron-density-weighted LOS magnetic field, and Zeeman splitting provides a more direct measurement of the LOS field strength \citep{Crutcher2012,Beck2015,Han2017}. These measurements are nevertheless limited by line-of-sight integration, Faraday depolarization, sensitivity, and the availability of suitable tracers \citep{Beck2015,Han2017}.

Magnetohydrodynamic (MHD) turbulence provides a complementary route. Theories of anisotropic MHD turbulence predict that turbulent eddies become increasingly elongated along the local magnetic field toward smaller scales \citep{GS95,LV99}. The anisotropy is caused by the fact that turbulence cascades preferentially perpendicular to the field rather than along it. Numerical simulations have established that this anisotropy is most clearly defined in the local magnetic-field frame and persists over a broad range of turbulent conditions \citep{2001ApJ...554.1175M,CL03,CV20,HXL21}. Consequently, the spatial gradients of turbulent fluctuations preferentially orient perpendicular to the local magnetic field. This geometrical relation provides the physical basis for gradient-based magnetic-field diagnostics.

The Gradient Technique was first developed using turbulent velocity fluctuations, leading to the Velocity Gradient Technique (VGT). Spectroscopic data provide two complementary velocity-sensitive observables: Velocity Centroid Gradients (VCGs), derived from velocity centroids \citep{GonzalezCasanova2017_160806867}, and Velocity Channel Gradients (VChGs), derived from thin velocity channels \citep{Lazarian2018_170303119}. Intensity Gradients (IGs) provide complementary information on density structures \citep{Hu2019_190809488}. The same principle extends to magnetic-field fluctuations through Synchrotron Intensity Gradients (SIGs) \citep{Lazarian2017_170107883} and Synchrotron Polarization Gradients (SPGs) \citep{Lazarian2018_180200028}. Spectroscopic and synchrotron gradients therefore probe different fluctuations with the same underlying turbulent anisotropy.

The standard gradient--magnetic-field alignment is most directly applicable to fluctuations dominated by the Alfv\'enic and slow-mode cascade. Intermittent fast modes, shocks, self-gravity, outflows, shear, and other coherent motions can compete with the turbulent velocity, density, and magnetic field, while thermal broadening and self-absorption alter how those fields are mapped into observations \citep{Ho2021_210408627,Hu2020_200206754,Hsieh2019_180909806,Ho2023_230113458}. These effects are physically informative: changes in gradient orientation or amplitude can identify transitions from turbulence-dominated gas to collapsing, shocked, or feedback-regulated structures.

This review develops a unified description of gradient techniques from theory to application. We first summarize the anisotropy of MHD turbulence and its implications for velocity, magnetic-field, and density gradients. We then discuss how turbulent anisotropy appears in spectroscopic and synchrotron observables, followed by a common formulation of VCGs, VChGs, IGs, SIGs, and SPGs. We finally review observational applications in the diffuse ISM, star-forming regions, the Galactic center, external galaxies, and galaxy clusters, with particular emphasis on magnetic-field tracing, tomography, and dynamical diagnostics.

\section{Anisotropy in MHD turbulence}
\label{sec:method}

\subsection{Scaling relations for turbulent velocity fluctuations}
\label{sec.2.2.1}
The fundamental theory of anisotropic incompressible MHD turbulence was
developed by \citet{GS95} for trans-Alfv\'enic turbulence,
$M_A\equiv v_{\rm inj}/v_A\approx1$, where $v_{\rm inj}$ is the turbulent
velocity at the injection scale $L_{\rm inj}$ and $v_A$ is the Alfv\'en
speed. \citet{GS95} introduced the ``critical balance'' condition, which
equates the nonlinear turbulent-cascade time with the Alfv\'en-wave
propagation time,
\begin{equation}
(k_\perp v_l)^{-1}\sim(k_\parallel v_A)^{-1},
\end{equation}
where $k_\parallel$ and $k_\perp$ are the wavevector components parallel and
perpendicular to the mean magnetic field, respectively, and $v_l$ is the
turbulent velocity associated with the perpendicular scale
$k_\perp^{-1}$.

Combining critical balance with the Kolmogorov-like perpendicular cascade,
\begin{equation}
v_l
\simeq
v_{\rm inj}(k_\perp L_{\rm inj})^{-1/3},
\end{equation}
gives the GS95 anisotropy scaling
\begin{equation}
\label{eq.gs95}
k_\parallel
=
(k_\perp L_{\rm inj})^{2/3}L_{\rm inj}^{-1},
\qquad
M_A\approx1.
\end{equation}
Thus, $k_\parallel$ is smaller than $k_\perp$ and turbulent eddies are anisotropically elongating along the magnetic field lines. However, the wavevectors are defined with respect to the mean magnetic field, corresponding to a global reference frame. In this frame, variations of the magnetic-field direction
on different spatial scales are averaged together, and the anisotropy of
large eddies can dominate over that of smaller eddies. As a result, the
intrinsic scale-dependent anisotropy is substantially weakened and can appear
approximately scale-independent when measured globally
\citep{CV20,2001ApJ...554.1175M,HXL21}.

\begin{figure*}
\centering
\includegraphics[width=.99\linewidth]{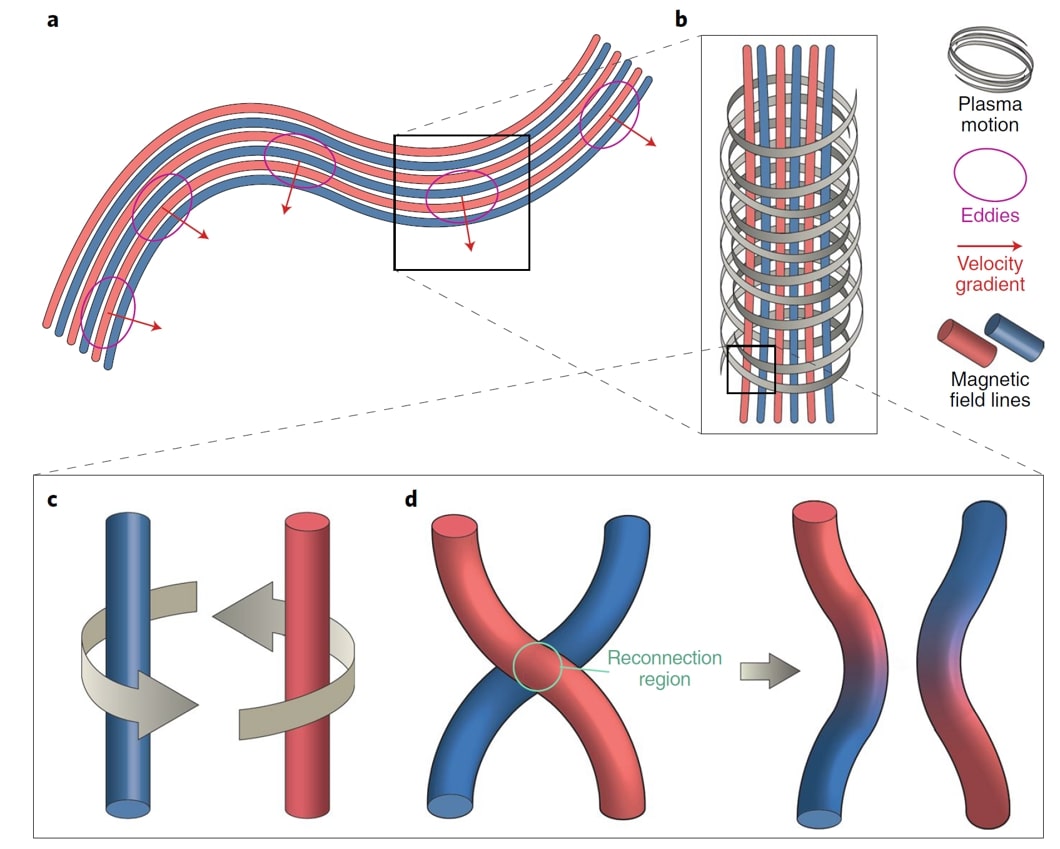}
\caption{
Schematic illustration of the local anisotropy of magnetohydrodynamic (MHD) turbulence, adapted from \citet{2019NatAs...3..692P}. Alfv\'enic turbulent eddies become elongated along the local magnetic field, while turbulent mixing proceeds preferentially across the field. As a result, the strongest local velocity variation occurs perpendicular to the magnetic-field direction, producing velocity gradients that are preferentially perpendicular to the local field.}
\label{fig:cartoon}
\end{figure*}

\citet{LV99} introduced the local reference frame, in which turbulent eddies are described relative to the magnetic field passing through the eddy at the scale under consideration. A key element of this picture is fast turbulent reconnection, which allows magnetic-field lines to change connectivity on the eddy-turnover timescale and thereby permits efficient mixing perpendicular to the local magnetic field (see Figure~\ref{fig:cartoon}). In this frame, the eddy dimensions $l_\perp$ and $l_\parallel$ are measured perpendicular and parallel to the
local field, and the strong-turbulence critical-balance condition becomes
\begin{equation}
\frac{v_{l,\perp}}{l_\perp}
\sim
\frac{v_A}{l_\parallel}.
\end{equation}
Numerical simulations subsequently confirmed that the scale-dependent
anisotropy is recovered when turbulent fluctuations are measured relative to
the local rather than the global magnetic field
\citep{CV20,2001ApJ...554.1175M,HXL21}.

\begin{figure*}[htbp!]
\centering
\includegraphics[width=.99\linewidth]{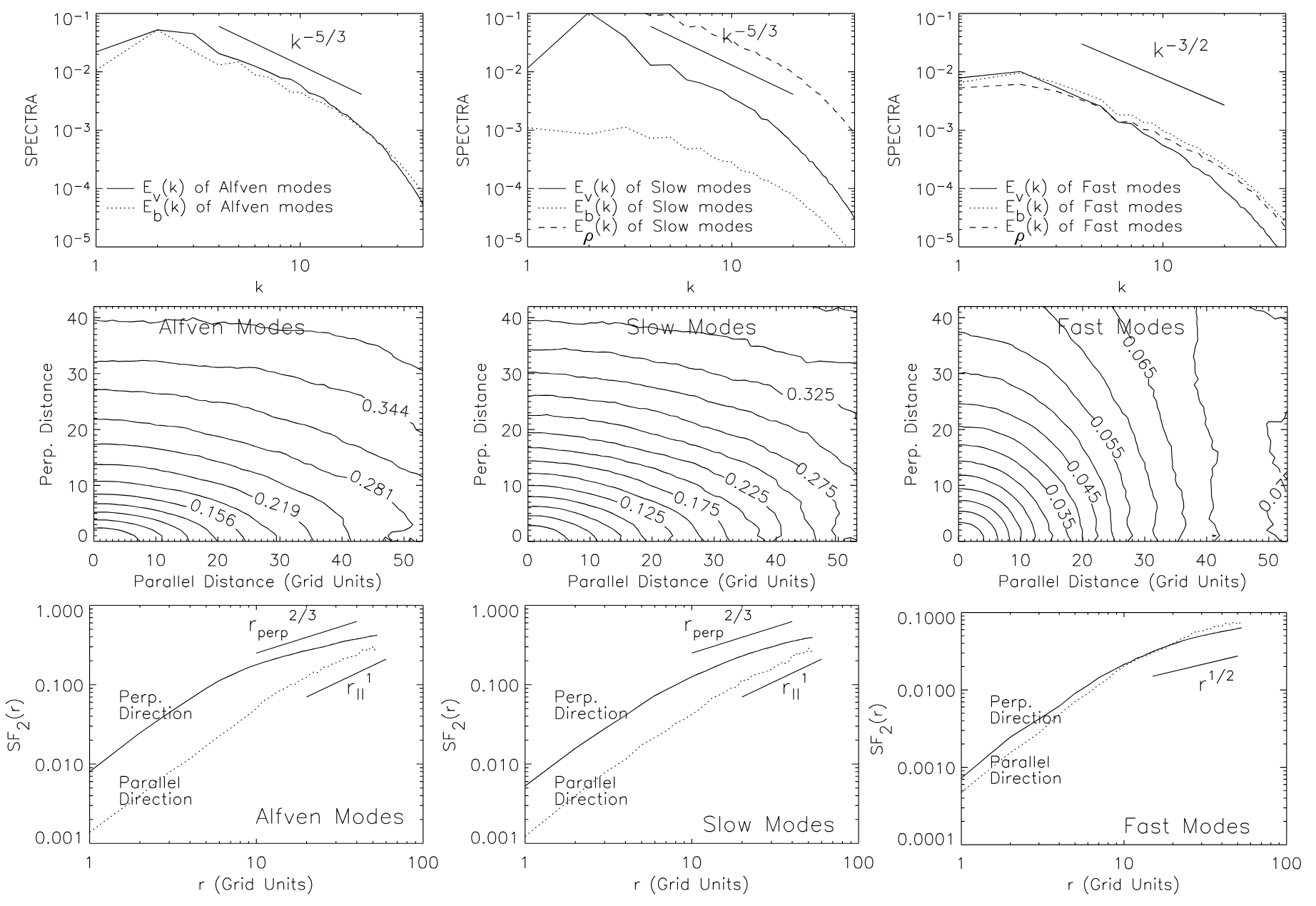}
\caption{Scaling and anisotropy of the three MHD turbulent modes in low-$\beta$ turbulence, adapted from \citet{CL03}. The simulation has $\beta\simeq0.2$, $M_s\simeq2.3$, and $M_A\simeq0.7$. Columns correspond to Alfv\'en, slow, and fast modes, respectively. The upper panels show the velocity and magnetic-energy spectra: Alfv\'en and slow modes follow approximately Kolmogorov-like $k^{-5/3}$ scalings, whereas fast modes are closer to a $k^{-3/2}$ spectrum. The middle panels show contours of the second-order velocity structure function measured relative to the local magnetic field. Alfv\'en and slow modes exhibit strong scale-dependent anisotropy, consistent with $r_\parallel\propto r_\perp^{2/3}$, while fast modes are approximately isotropic. The lower panels compare the structure functions measured parallel and perpendicular to the local magnetic field, further demonstrating the anisotropic Alfv\'en and slow cascades and the nearly isotropic fast-mode cascade. 
}
\label{fig:CL03_modes}
\end{figure*}

The scale at which the strong anisotropic cascade develops depends on the Alfv\'en Mach number, $M_A=v_{\rm inj}/v_A$. Here, ``weak'' and ``strong'' refer to the strength of nonlinear interactions: weak turbulence with weak nonlinear interactions could be described as waves, whereas strong turbulence satisfies the critical-balance condition. For super-Alfv\'enic turbulence ($M_A>1$), the magnetic field is dynamically weak at the injection scale and the turbulence is initially approximately hydrodynamic and isotropic. As the cascade proceeds to smaller scales, the turbulent velocity decreases until magnetic backreaction becomes important at the Alfv\'en scale
\begin{equation}
l_A=L_{\rm inj}M_A^{-3},
\end{equation}
below which the cascade enters the strong anisotropic MHD regime. For sub-Alfv\'enic turbulence ($M_A<1$), magnetic tension is dynamically important already at the injection scale, but the nonlinear interactions are initially weak. The cascade proceeds predominantly toward smaller perpendicular scales and becomes progressively stronger until critical balance is reached at the transition scale
\begin{equation}
l_{\rm trans}=L_{\rm inj}M_A^2.
\end{equation}
Below $l_{\rm trans}$, the turbulence enters the strong anisotropic regime. In the strong-turbulence regime, the local anisotropy is given by
\begin{equation}
\label{eq.lv99_l}
l_\parallel=
\begin{cases}
L_{\rm inj}
\left(\dfrac{l_\perp}{L_{\rm inj}}\right)^{2/3}
M_A^{-1},
& M_A>1,\quad l_\perp<l_A,\\[7pt]
L_{\rm inj}
\left(\dfrac{l_\perp}{L_{\rm inj}}\right)^{2/3}
M_A^{-4/3},
& M_A<1,\quad l_\perp<l_{\rm trans},
\end{cases}
\end{equation}
with the corresponding perpendicular velocity scaling
\begin{equation}
\label{eq.lv99_v}
v_{l,\perp}=
\begin{cases}
v_{\rm inj}
\left(\dfrac{l_\perp}{L_{\rm inj}}\right)^{1/3},
& M_A>1,\quad l_\perp<l_A,\\[7pt]
v_{\rm inj}
\left(\dfrac{l_\perp}{L_{\rm inj}}\right)^{1/3}
M_A^{1/3},
& M_A<1,\quad l_\perp<l_{\rm trans}.
\end{cases}
\end{equation}
These predictions have been extensively tested in numerical simulations (see Figure~\ref{fig:CL03_modes} and \citep{2001ApJ...554.1175M,CL03,2010ApJ...720..742K,CV20,HXL21,2024MNRAS.527.3945H}) and are broadly consistent with in situ measurements of solar-wind turbulence \citep{2016ApJ...816...15W,2020FrASS...7...83M,2021ApJ...915L...8D,2023arXiv230512507Z}.

Equation~\ref{eq.lv99_l} shows that turbulent eddies become increasingly elongated along the local magnetic field toward smaller scales. Since $l_\parallel>l_\perp$, turbulent fluctuations vary more rapidly across the local magnetic field than along it. The velocity gradient is therefore dominated by its perpendicular component,
\begin{equation}
\label{eq.vg}
\nabla v_l
\approx
\frac{v_{l,\perp}}{l_\perp}\hat{\boldsymbol{l}}_\perp
=
\begin{cases}
\dfrac{v_{\rm inj}}{L_{\rm inj}}
\left(\dfrac{l_\perp}{L_{\rm inj}}\right)^{-2/3}
\hat{\boldsymbol{l}}_\perp,
& M_A>1,\\[8pt]
\dfrac{v_{\rm inj}}{L_{\rm inj}}
\left(\dfrac{l_\perp}{L_{\rm inj}}\right)^{-2/3}
M_A^{1/3}\hat{\boldsymbol{l}}_\perp,
& M_A<1.
\end{cases}
\end{equation}
Thus, in the strong Alfv\'enic-turbulence regime, velocity gradients are preferentially perpendicular to the local magnetic field (see Fig.~\ref{fig:cartoon}).

\subsection{Scaling relations for turbulent magnetic-field and density fluctuations}
The anisotropy of MHD turbulence is also imprinted on magnetic-field fluctuations. For Alfv\'enic perturbations,
\begin{equation}
\frac{\delta B_l}{B_0}
\sim
\frac{v_{l,\perp}}{v_A},
\label{eq:alfven_B_relation}
\end{equation}
so magnetic-field fluctuations inherit the anisotropy of the Alfv\'enic velocity field and are dominated by variations perpendicular to the local magnetic field \citep{CL02,CL03}. Their gradients are therefore primarily determined by variations across the local field,
\begin{equation}
\nabla\delta B_l
\sim
\frac{\delta B_l}{l_\perp}\hat{\boldsymbol{l}}_\perp
\propto
l_\perp^{-2/3}\hat{\boldsymbol{l}}_\perp,
\label{eq:B_gradient}
\end{equation}
and are consequently preferentially perpendicular to the local magnetic-field direction.

The extension to density fluctuations requires more care because the Alfv\'en mode itself is incompressible. Mode-decomposition studies of compressible MHD turbulence show that slow modes are mixed by the Alfv\'enic cascade and approximately inherit its scale-dependent anisotropy, whereas fast modes are substantially more isotropic \citep{CL02,CL03}. Density fluctuations dominated by the Alfv\'enic/slow-mode cascade can therefore exhibit the same preferred local orientation,
\begin{equation}
\nabla\delta\rho_l
\sim
\frac{\delta\rho_l}{l_\perp}\hat{\boldsymbol{l}}_\perp.
\label{eq:rho_gradient}
\end{equation}

More generally, if a turbulence-dominated fluctuation follows $\delta X_l\propto l_\perp^\alpha$, its gradient amplitude scales as
\begin{equation}
|\nabla X_l|
\sim
\frac{\delta X_l}{l_\perp}
\propto
l_\perp^{\alpha-1}.
\label{eq:general_gradient_scaling}
\end{equation}
Kolmogorov-like scaling, $\alpha=1/3$, gives $|\nabla X_l|\propto l_\perp^{-2/3}$, whereas a Burgers-like amplitude scaling, $\alpha=1/2$, gives $|\nabla X_l|\propto l_\perp^{-1/2}$. In either case, turbulence-generated gradients become stronger toward smaller scales \citep{Hu2025_241008157}. 

\subsection{Departures from the gradient and magnetic field alignment}
\label{sec:gradient_departures}

The relations above describe fluctuations governed primarily by the Alfv\'enic/slow-mode cascade. Other turbulent modes, dynamical processes, and radiative-transfer effects may produce systematic departures from the standard perpendicular-gradient relation.

Fast modes provide an important turbulent exception. Although their cascade is substantially more isotropic on average \citep{CL02,CL03}, fast modes can be highly intermittent in magnetically dominated, low-$\beta$ turbulence (see Figure~\ref{fig:CL03_modes}). \citet{Ho2021_210408627} showed that locally fast-mode-dominated regions can produce anomalous gradient orientations, including approximately $90^\circ$ changes relative to the Alfv\'enic alignment, even when the volume-averaged fast-mode energy is subdominant. Numerical studies of subsonic turbulence further showed that magnetic-field tracing remains effective over a broad range of magnetization, while fast-mode intermittency introduces localized deviations \citep{Ho2023_230113458}.

Shocks and self-gravity alter the underlying dynamics more directly. Across a shock, the strongest velocity and density variations are set by the shock normal rather than by local eddy anisotropy, and numerical studies show corresponding changes in gradient amplitudes and relative orientations \citep{KaHoYuen2017_170303026,GrandaMunoz2025_240508702}. Self-gravity produces coherent converging motions and can drive an approximately $90^\circ$ reorientation of velocity gradients relative to the turbulence-dominated state \citep{KaHoYuen2017_170303026,Hu2020_200206754}. Changes in gradient amplitude, orientation statistics, and curvature can therefore provide diagnostics of the onset and evolution of gravitational collapse. Other coherent motions, including stellar outflows, expanding shells, and large-scale shear, can similarly impose gradients determined by the geometry of the organized flow.

Radiative-transfer effects should be distinguished from these dynamical departures because they modify how the underlying turbulent velocity and density fields are mapped into observed spectral-line emission without necessarily altering the gas dynamics. In molecular clouds, line intensities depend on excitation conditions, molecular abundances, and optical depth, with self-absorption progressively suppressing emission from deeper regions along the line of sight \citep{LazarianPogosyan2004,Roueff2021}. Different molecular tracers therefore sample distinct cloud environments: optically thick $^{12}$CO emission is generally dominated by extended molecular gas and foreground layers near the $\tau\sim1$ surface, while the less abundant $^{13}$CO and C$^{18}$O isotopologues can probe more deeply into shielded, higher-column-density regions. These differences are further influenced by selective photodissociation, chemical fractionation, and molecular depletion \citep{Roueff2021,Tafalla2023}. Despite these complications, radiative-transfer calculations demonstrate that velocity centroid gradients retain their ability to trace magnetic-field orientations over a broad range of optical depths \citep{GonzalezCasanova2019,Hsieh2019_180909806}. Complementary sampling of CO isotopologues also probes magnetic-field structure associated with different emitting regions along the line of sight \citep{Hsieh2019_180909806,Hu2021_200105648}. Simulations incorporating self-gravity and self-absorption further show that gradient orientations remain comparatively robust, whereas optical-depth effects can generate power-law tails in intensity PDFs that mimic signatures of gravitational collapse \citep{Hu2021_200105648}.

Thermal broadening further modifies the observed velocity-channel structures by redistributing emission over a finite velocity range, potentially reducing the contrast of small-scale velocity caustics, particularly in warm gas. However, this smoothing does not eliminate the overall contribution of velocity caustics. Using multiphase H I simulations that include thermal broadening and GALFA-H I observations, \citet{2023MNRAS.tmp.1894H} demonstrated that velocity fluctuations remain the dominant contributor to intensity structures in thin velocity channels (see also Figure~\ref{fig:vc}), confirming that the observed H I striations arise predominantly from velocity caustics rather than density fluctuations.

These effects fall into three physically distinct categories: changes in turbulent mode, competition with non-turbulent dynamical processes, and observational transfer effects. Distinguishing among them is essential when gradient orientations are used to infer either magnetic-field structure or local gas dynamics.

\section{Observational manifestations of MHD turbulence anisotropy}

\subsection{Density and velocity fluctuations in spectroscopic observations}
\label{sec:spec_anisotropy}

Spectroscopic observations provide a position--position--velocity (PPV) cube, $\rho_s(x,y,v)$, that mixes information about the spatial distribution of the emitting gas with its line-of-sight (LOS) velocity. For an optically thin line,
\begin{equation}
\label{eq.max}
\rho_s(x,y,v)
=
\kappa\int dz\,
\rho(\pmb{x})
\phi\!\left[v-v_{\rm tur}(\pmb{x})-v_{\rm gal}(\pmb{x})\right],
\end{equation}
where $\rho(\pmb{x})$ is the gas density, $v_{\rm tur}$ is the turbulent LOS velocity, $v_{\rm gal}$ represents coherent motions such as Galactic rotation, and $\phi$ accounts for thermal broadening. A channel map centered at $v_0$ with width $\Delta v$ is
\begin{equation}
p(x,y,v_0)
=
\int_{v_0-\Delta v/2}^{v_0+\Delta v/2}
\rho_s(x,y,v)\,dv.
\label{eq:channel_map}
\end{equation}

\subsubsection{Density fluctuations}
Different projections of the PPV cube emphasize different turbulent quantities. The velocity-integrated intensity, or moment-0 map,
\begin{equation}
M_0(x,y)=\int \rho_s(x,y,v)\,dv,
\label{eq:moment0}
\end{equation}
is proportional to the column density of the emitting species for an optically thin line with approximately uniform abundance and excitation. 
Opacity, excitation, chemistry, and multiphase structure can modify the correspondence between integrated intensity and the underlying density field.

\subsubsection{Velocity fluctuations}
Velocity fluctuations can be accessed through both velocity centroids and thin velocity channels. The velocity centroids, or moment-1 map,
\begin{equation}
M_1(x,y)
=
\frac{\int v\,\rho_s(x,y,v)\,dv}
{\int \rho_s(x,y,v)\,dv},
\label{eq:moment1}
\end{equation}
provides a projected, density-weighted measure of the LOS velocity. Centroid statistics recover the scaling and anisotropy of the turbulent velocity field under appropriate conditions, although strong density fluctuations can contaminate highly compressible turbulence \citep{2003ApJ...592L..37L,2005ApJ...631..320E,2014ApJ...790..130B,2020ApJ...901...11H}.

Velocity fluctuations can also be obtained from a velocity channel $p(x,y,v_0)$. Following \citet{LP00,LP04}, whether $p(x,y,v_0)$ is ``thin'' or ``thick'' depends on its effective velocity width relative to the turbulent velocity dispersion at the spatial scale of interest. A channel is considered thin when
\begin{equation}
\Delta v_{\rm eff} \lesssim \sigma_v(l),
\end{equation}
and thick when
\begin{equation}
\Delta v_{\rm eff} \gtrsim \sigma_v(l),
\end{equation}
where $\Delta v_{\rm eff}$ includes both the instrumental channel width and thermal broadening, and $\sigma_v(l)$ is the turbulent velocity dispersion at scale $l$. 

\begin{figure*}[htbp!]
\includegraphics[width=1.0\linewidth]{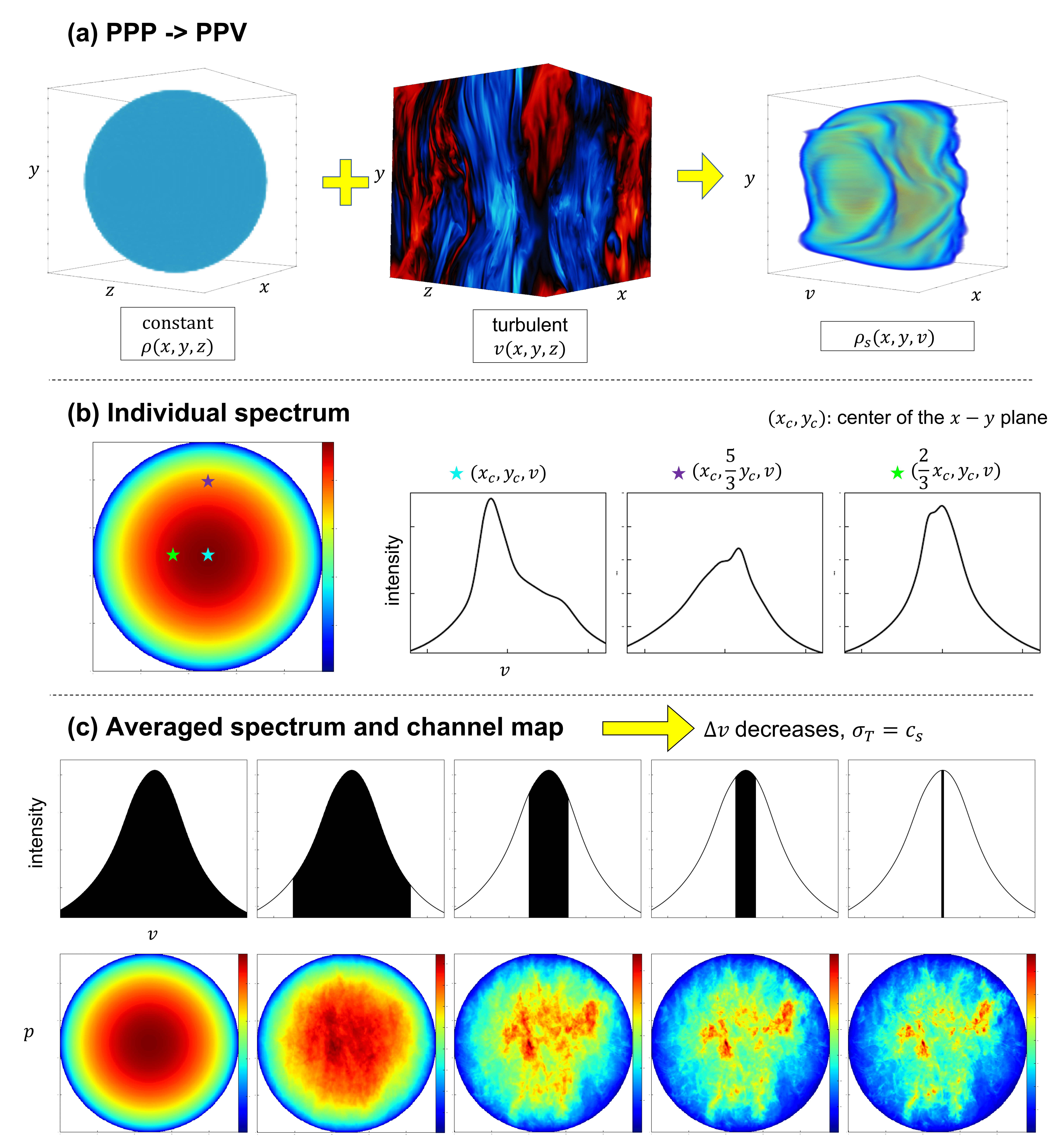}
\caption{Numerical illustration of velocity caustics. Panel (a) shows a synthetic PPV cube $\rho_s(x,y,v)$ generated from a spatially uniform density distribution and the LOS velocity field of a multiphase turbulence simulation. Panel (b) shows representative spectra at three positions in the projected map. Panel (c) shows the spatially averaged velocity spectrum and a representative velocity-channel map. The shaded region indicates the velocity interval used to construct the channel. Complex spatial structures appear despite the absence of internal density fluctuations, illustrating turbulent velocity mapping. Adapted from Hu et al. \citep{2023MNRAS.tmp.1894H}.}
\label{fig:vc}
\end{figure*}
Thin velocity channels provide a complementary velocity-sensitive diagnostic through PPV mapping. Spatially separated gas parcels with similar LOS velocities can overlap within the same channel, whereas different portions of a single physical structure can be separated in velocity space. This velocity crowding, commonly referred to as ``velocity caustics,'' allows turbulent velocity fluctuations to generate small-scale structures in channel maps even in the absence of corresponding density fluctuations \citep{LP00,LP04}. Figure~\ref{fig:vc} illustrates this effect using a spatially uniform density field mapped into PPV space by a turbulent velocity field.

The relative contributions of density and velocity depend systematically on the channel width. Velocity fluctuations become increasingly important in thin channels, whereas increasing the channel width progressively averages over velocity fluctuations and shifts the channel statistics toward those of the projected density field \citep{LP00,LP04}. Numerical and observational studies of atomic, molecular, and ionized gas have confirmed this thin--thick transition and demonstrated that turbulent velocity statistics can be recovered from PPV observations \citep{2003MNRAS.342..325E,2009ApJ...693.1074C,2006ApJ...653L.125P,2008ApJS..174..202S,2016MNRAS.463.2864A}. Synthetic PPV studies further showed that the anisotropy of thin-channel fluctuations retains information about both the magnetic-field orientation and the magnetization of the turbulence \citep{2003MNRAS.342..325E,2015ApJ...814...77E,2016MNRAS.461.1227K}.

Velocity centroids and thin velocity channels therefore provide complementary projections of the turbulent velocity field. The moment-1 map, $M_1$, represents an LOS-integrated, intensity-weighted velocity measure, whereas thin channels retain velocity-resolved spatial structures generated through PPV mapping.

\subsection{Magnetic-field fluctuations in synchrotron intensity and polarization}
\label{sec:synchrotron_anisotropy}
Synchrotron emission provides a direct observational probe of turbulent magnetic-field fluctuations. For a power-law relativistic-electron distribution, $N(E)\propto E^{-p}$, the synchrotron intensity can be written schematically as
\begin{equation}
I_{\rm syn}(\pmb{X})
\propto
\int dz\,n_{\rm cr}(\pmb{x})B_\perp^{\gamma_{\rm syn}}(\pmb{x}),
\qquad
\gamma_{\rm syn}=\frac{p+1}{2},
\label{eq:synch_intensity}
\end{equation}
where $\pmb{X}$ is the POS coordinate, $n_{\rm cr}$ is the relativistic-electron density, and $B_\perp$ is the magnetic-field component perpendicular to the LOS. In addition to magnetic-field fluctuations, spatial variations in $n_{\rm cr}$ may also contribute to synchrotron-intensity fluctuations and can be correlated with the magnetic field through cosmic-ray transport. Under the common approximation that magnetic fluctuations dominate, synchrotron emission retains information about the MHD turbulence's anisotropy \citep{LP12}.

Synchrotron polarization additionally encodes field orientation and Faraday rotation. Defining the complex linear polarization
\begin{equation}
\mathcal{P}(\pmb{X},\lambda^2)=Q+iU,
\end{equation}
where $Q$ and $U$ are the Stokes parameters describing linear polarization along two orthogonal reference axes and axes rotated by $45^\circ$, respectively, the polarized emission can be written schematically as
\begin{equation}
\mathcal{P}(\pmb{X},\lambda^2)
\propto
\int dz\,
n_{\rm cr}B_\perp^{\gamma_{\rm syn}}
\exp\left\{2i\left[\psi_0(\pmb{x})+\lambda^2\Phi(\pmb{x})\right]\right\}.
\label{eq:synch_pol}
\end{equation}
Here, $\psi_0$ is the intrinsic polarization angle and $\Phi$ is the Faraday depth between the emitting location and the observer. The polarized intensity is $P=|\mathcal{P}|=\sqrt{Q^2+U^2}$, while the polarization angle is $\psi=\frac12\operatorname{atan2}(U,Q)$. At short wavelengths, polarized synchrotron fluctuations largely inherit the anisotropy of the emitting magnetic field; at longer wavelengths, Faraday rotation and depolarization introduce additional sensitivity to the thermal electron density and LOS magnetic field.

Multi-frequency polarization theory shows that the wavelength dependence of these statistics can separate properties of the synchrotron-emitting turbulence from those of the Faraday-rotating medium \citep{LP16}. Numerical studies confirm that synchrotron intensity and polarization remain anisotropic across a broad range of magnetization, viewing angles, wavelengths, and MHD-mode mixtures, with Alfv\'en and slow modes exhibiting stronger anisotropy than fast modes \citep{2019ApJ...877..108L,2020ApJ...890...70W,CL03}.

\section{Calculation of gradient techniques}
\label{sec:gradient_methods}

The Velocity Gradient Technique (VGT), Synchrotron Intensity Gradient (SIG), and Synchrotron Polarization Gradient (SPG) techniques share a common procedure for extracting the preferred orientation of turbulent fluctuations \citep{GonzalezCasanova2017_160806867,Lazarian2017_170107883,Lazarian2018_180200028}. Their principal difference is the observable from which the spatial gradient is calculated.

For a two-dimensional observable $f(x,y)$, the gradient components are evaluated as
\begin{equation}
\begin{aligned}
\nabla_x f(x,y) &= G_x*f(x,y),\\
\nabla_y f(x,y) &= G_y*f(x,y),
\end{aligned}
\end{equation}
where $G_x$ and $G_y$ are Sobel kernels and $*$ denotes convolution. The pixelized gradient orientation is
\begin{equation}
\psi_g(x,y)
=
\mathrm{atan2}\!\left[\nabla_y f(x,y),\nabla_x f(x,y)\right].
\label{eq:gradient_angle}
\end{equation}

Because turbulent anisotropy is statistical, individual pixel-scale gradients provide noisy estimates of the preferred direction. The gradient map is therefore commonly processed with sub-block averaging \citep{GonzalezCasanova2017_160806867,Yuen2017_170107944}: the distribution of $\psi_g$ within each sub-block is constructed, and its peak defines the preferred orientation $\psi_{gs}$. A Gaussian smoothing step is then applied to improve spatial coherence. Because gradient orientations are axial quantities with a $180^\circ$ degeneracy, the smoothing can be performed on
\begin{equation}
q_g=f\cos(2\psi_{gs}),\qquad
u_g=f\sin(2\psi_{gs}),
\end{equation}
so that
\begin{equation}
\widetilde{q}_g=G_\sigma*q_g,\qquad
\widetilde{u}_g=G_\sigma*u_g,
\end{equation}
where $G_\sigma$ is a two-dimensional Gaussian kernel. The smoothed orientation is
\begin{equation}
\widetilde{\psi}_{gs}
=
\frac{1}{2}
\mathrm{atan2}\!\left(\widetilde{u}_g,\widetilde{q}_g\right).
\end{equation}
For fluctuations dominated by the Alfv\'enic/slow-mode cascade, the inferred POS magnetic-field orientation is
\begin{equation}
\psi_B=\widetilde{\psi}_{gs}+\frac{\pi}{2}.
\label{eq:gradient_B}
\end{equation}
The same gradient, sub-block averaging, and smoothing procedure is applied to VChGs, VCGs, IGs, SIGs, and SPGs. Noisy data points may need to be removed before the smoothing.

\subsection{Velocity Gradient Technique}
\label{sec:vgt_method}

The VGT uses velocity-sensitive spectroscopic observables \citep{GonzalezCasanova2017_160806867,Yuen2017_170107944,Lazarian2018_170303119}. For Velocity Channel Gradients (VChGs),
\begin{equation}
f(x,y)=p(x,y,v_0),
\end{equation}
where $p$ is the intensity in a thin velocity channel. When multiple channels are combined, the sub-block-averaged orientations are integrated through pseudo-Stokes parameters \citep{Hu2022_220605423,Liu2022_220806090},
\begin{align}
Q_g(x,y)
&=
\int p(x,y,v_0)
\cos\!\left[2\psi_{gs}(x,y,v_0)\right]\,dv_0,\\
U_g(x,y)
&=
\int p(x,y,v_0)
\sin\!\left[2\psi_{gs}(x,y,v_0)\right]\,dv_0.
\label{eq:vgt_pseudostokes}
\end{align}
The pseudo-Stokes maps may be Gaussian smoothed before recovering the combined orientation,
\begin{equation}
\psi_g(x,y)
=
\frac{1}{2}\mathrm{atan2}\!\left[U_g(x,y),Q_g(x,y)\right],
\end{equation}
followed by the $90^\circ$ rotation in Eq.~\ref{eq:gradient_B}. The quantities $Q_g$ and $U_g$ are mathematical constructs for combining gradient orientations and should not be confused with physical polarization Stokes parameters.

Velocity Centroid Gradients (VCGs) use
\begin{equation}
f(x,y)=M_1(x,y),
\end{equation}
where $M_1$ is defined in Eq.~\ref{eq:moment1}. Because $M_1$ is intensity-weighted, strong density fluctuations can modify centroid statistics in highly compressible turbulence \citep{2005ApJ...631..320E}. 

Intensity Gradients (IGs) instead use
\begin{equation}
f(x,y)=M_0(x,y),
\end{equation}
or a thick velocity channel, and therefore respond primarily to projected density structure. Density, however, is not a clean tracer of the turbulent cascade, as its fluctuations can be strongly modified by compressibility, shocks, gravity, and other dynamical processes; even the density spectral slope depends sensitively on the sonic Mach number $M_s$ \citep{KimRyu2005,Kowal2007,FederrathKlessen2013}. 

\subsection{Synchrotron Intensity Gradients}
\label{sec:sig_method}

SIGs apply the common gradient procedure to
\begin{equation}
f(x,y)=I_{\rm syn}(x,y),
\end{equation}
where $I_{\rm syn}$ is defined in Eq.~\ref{eq:synch_intensity} \citep{Lazarian2017_170107883}. SIGs therefore probe the anisotropy of magnetic-field fluctuations directly from synchrotron total intensity.

\subsection{Synchrotron Polarization Gradients}
\label{sec:spg_method}
SPGs apply the same procedure to the polarization intensity,
\begin{equation}
f(x,y)=P(x,y,\lambda^2)=|\mathcal{P}(x,y,\lambda^2)|,
\end{equation}
with $\mathcal{P}$ defined in Eq.~\ref{eq:synch_pol} \citep{Lazarian2018_180200028}. Because $\mathcal{P}$ depends on Faraday rotation, the structures sampled by SPGs vary with observing wavelength. In the strong-Faraday-rotation regime, wavelength-dependent depolarization restricts the effective LOS depth contributing to the observed polarization, allowing different frequencies to probe different portions of the magnetized medium \citep{LP16,Zhang2019_190411939}.

The input observables and the turbulent fluctuations they primarily trace can therefore be summarized as
\begin{equation}
\begin{aligned}
{\rm VChG}:&\quad p(x,y,v_0)
&&\rightarrow\ {\rm velocity\ fluctuations},\\
{\rm VCG}:&\quad M_1(x,y)
&&\rightarrow\ {\rm velocity\ fluctuations},\\
{\rm IG}:&\quad M_0(x,y)
&&\rightarrow\ {\rm density\ fluctuations},\\
{\rm SIG}:&\quad I_{\rm syn}(x,y)
&&\rightarrow\ {\rm magnetic\text{-}field\ fluctuations},\\
{\rm SPG}:&\quad P(x,y,\lambda^2)
&&\rightarrow\ {\rm magnetic\text{-}field\ fluctuations}.
\end{aligned}
\end{equation}
The techniques share the same gradient calculation and statistical averaging but probe different turbulent fluctuations. VChGs retain velocity-resolved PPV information, while SPGs additionally retain frequency-dependent information associated with synchrotron polarization and Faraday rotation.

\begin{figure*}[htbp!]
\includegraphics[width=1.0\linewidth]{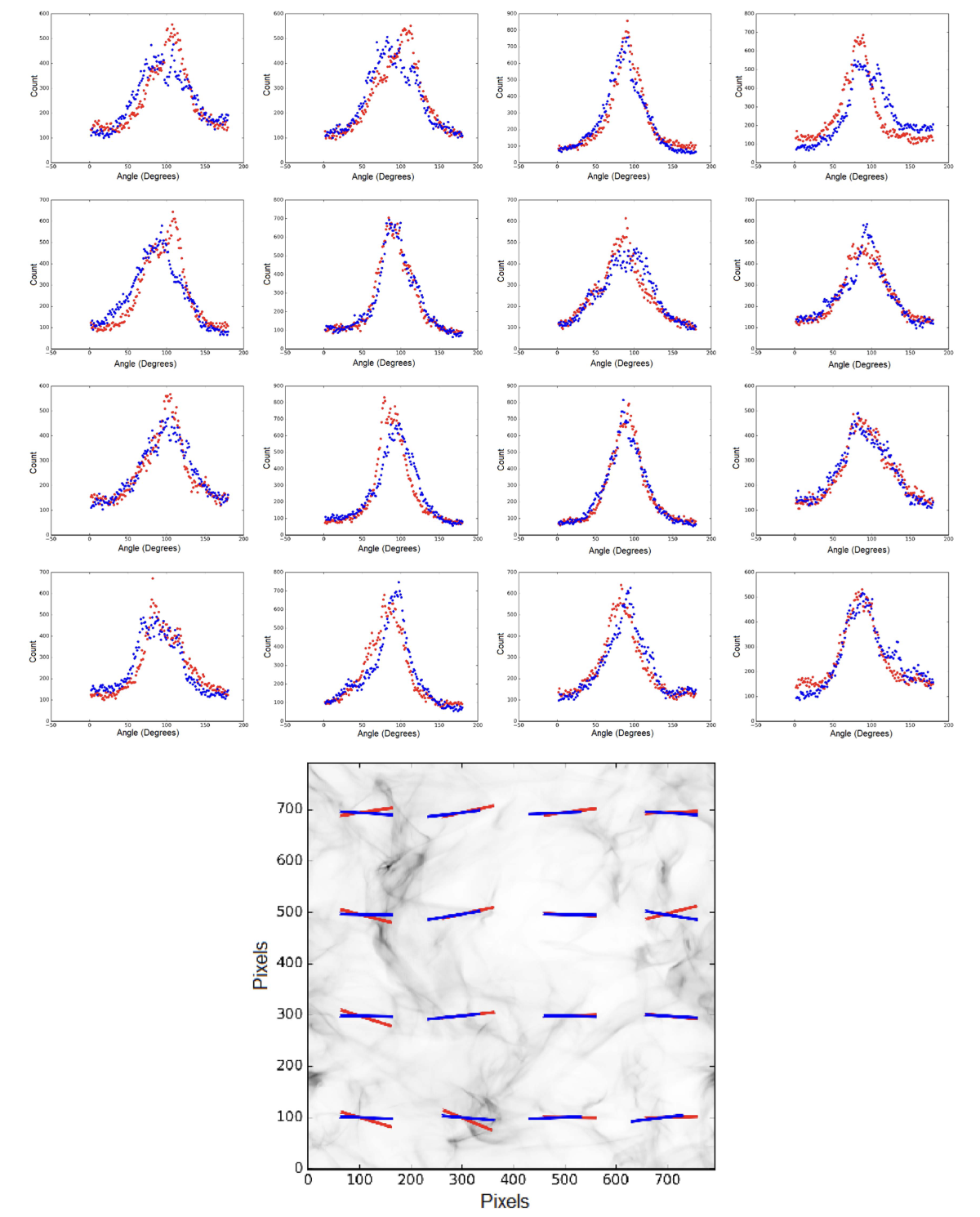}
\caption{Illustration of the sub-block averaging procedure introduced by
\citet{Yuen2017_170107944}. The upper panels show the distributions of the
VCGs' orientation (red) and the relative angle between the
VCGs and polarization (blue) in 16 sub-blocks of a synthetic
velocity centroid map. Although individual gradients exhibit substantial scatter, each sub-block develops a statistically preferred orientation; the peak of the gradient-angle distribution is used to determine the mean gradient direction. The lower panel compares the gradient directions recovered from sub-block averaging (red) with the synthetic polarization in the simulation (blue).
}
\label{fig:subblock}
\end{figure*}

\subsection{Techniques for improving gradient accuracy}

The accuracy of gradient techniques can be further improved by isolating the turbulent fluctuations that most directly reflect the local magnetic field. Several complementary approaches have been developed for this purpose.

The most important step is sub-block averaging, because turbulent anisotropy is intrinsically a statistical property rather than a point-by-point relation. Individual pixel-scale gradients therefore need not be exactly perpendicular to the local magnetic field and naturally exhibit substantial scatter, further broadened by intermittency and observational or numerical noise. As shown in Figure~\ref{fig:subblock}, by sampling the distribution of gradient orientations within a finite region, sub-block averaging recovers the statistically preferred direction associated with the local turbulent anisotropy \citep{Yuen2017_170107944}. The choice of sub-block size thus represents a trade-off between statistical accuracy and spatial resolution. Analyses that omit this statistical averaging and compare gradients with magnetic-field orientations pixel by pixel can consequently report substantially lower alignment accuracy.

Principal Component Analysis (PCA) can be applied to spectroscopic PPV data before calculating velocity gradients \citep{Hu2018_180208772}. By decomposing the PPV cube into principal components, PCA preferentially retains coherent spectroscopic fluctuations while suppressing redundant or noisy information. In practice, it acts as a form of filtering along the velocity dimension and has been shown to improve the alignment between velocity gradients and magnetic-field
orientations in both synthetic observations and Galactic H\,I data.

Spatial filtering in Fourier space provides another way to isolate the anisotropic turbulent cascade \citep{2024ApJ...972...17H}. Large-scale fluctuations can be weakly anisotropic or dominated by non-turbulent motions and therefore degrade gradient alignment. Removing low-$k$ modes emphasizes the smaller-scale turbulent fluctuations whose gradients more closely follow the local magnetic field. This is particularly useful in super-Alfv\'enic turbulence, where scales larger than $l_A$ are approximately isotropic and magnetic-field tracing becomes effective only below the Alfv\'en scale.

For velocity-channel gradients, the Velocity Decomposition Algorithm (VDA) provides a more direct separation of the PPV fluctuations associated with velocity and density \citep{Yuen2021_201215776}. VDA decomposes a channel map into velocity-dominated and density-dominated components, allowing gradients to be calculated from the velocity contribution alone. This reduces contamination from density structures, particularly in compressible and multiphase media, and improves the recovery of magnetic-field orientations from thin velocity channels.

These methods address different sources of contamination and can therefore be
combined: PCA selects the most informative spectroscopic components,
$k$-space filtering restricts the analysis to the relevant turbulent scales,
and VDA isolates the velocity contribution within individual channel maps.

\section{Observational applications}
\subsection{Magnetic field in diffuse ISM}
\label{sec:applications_diffuseISM}

\begin{figure*}[htbp!]
\includegraphics[width=1.0\linewidth]{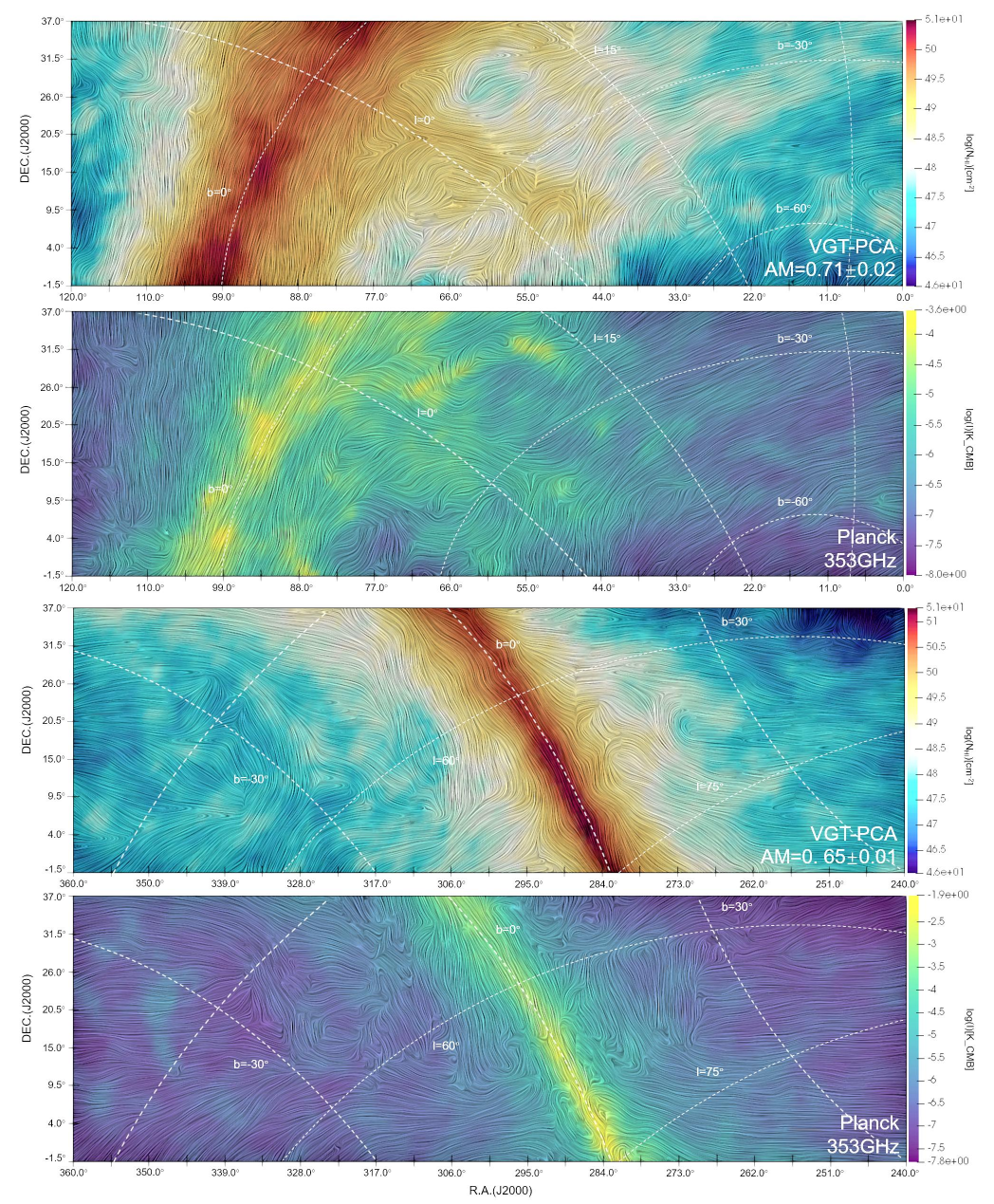}
\caption{Comparison of the plane-of-sky magnetic-field morphology inferred from the
PCA-enhanced Velocity Gradient Technique (VGT-PCA) with that traced by Planck 353~GHz dust polarization in two large regions of the Galactic sky. The first and third panels show the magnetic-field orientations reconstructed with VGT-PCA, while the second and fourth panels show the corresponding Planck polarization-derived magnetic-field orientations. Streamlines indicate the inferred POS magnetic-field direction, and the dashed curves mark Galactic longitude and latitude. The close morphological agreement demonstrates that velocity gradients statistically recover the large-scale and local magnetic-field structure over extended diffuse H\,{\sc i} regions. The corresponding alignment measures are ${\rm AM}=0.71\pm0.02$ and $0.65\pm0.01$ for the two regions, respectively. Adapted from Hu et al. \cite{2020ApJ...888...96H}.}
\label{fig:HI}
\end{figure*}

The spectroscopic development began with H\,{\sc i} observations. VCGs derived from GALFA-H\,{\sc i} recovered POS magnetic-field orientations consistent with Planck 353~GHz ($\lambda\simeq850\,\mu{\rm m}$) thermal-dust polarization, establishing that turbulent velocity anisotropy is measurable in diffuse atomic gas \citep{Yuen2017_170107944}. VChGs extended this approach to thin velocity channels, allowing magnetic structures associated with different LOS velocities to be separated rather than fully integrated in projection \citep{Lazarian2018_170303119}. The capability of VChGs to recover the POS magnetic-field morphology is illustrated in Figure~\ref{fig:HI}, where the PCA-enhanced VChG reconstruction closely follows the magnetic-field orientation inferred independently from Planck 353~GHz dust polarization over large regions of the diffuse H\,{\sc i} sky \citep{2020ApJ...888...96H}. The agreement is quantified by the alignment measure
\begin{equation}
{\rm AM}=2\left\langle\cos^2\theta\right\rangle-1
=\left\langle\cos 2\theta\right\rangle,
\end{equation}
where $\theta$ is the relative angle between the magnetic-field orientations inferred from the gradient technique and polarization. ${\rm AM}=1$ corresponds to perfect alignment, ${\rm AM}=0$ to random relative orientations, and
${\rm AM}=-1$ to perpendicular orientations. The two regions shown in Figure~\ref{fig:HI} yield ${\rm AM}=0.71$ and $0.65$, demonstrating the statistical accuracy achievable after PCA filtering and sub-block averaging.

A major advantage of VChGs is their ability to separate magnetic-field structures associated with different velocity components along the LOS, addressing the foreground and background contamination inherent in integrated polarization measurements. While dust polarization generally combines emission from multiple gas structures along the LOS, VChGs use spectroscopic velocity information to distinguish their magnetic-field orientations. When combined with velocity--distance information, this capability further enables distance-resolved magnetic-field tomography. The approach was first demonstrated using Galactic H\,{\sc i} observations and validated against optical starlight polarization toward stars with independently measured distances, showing good agreement between the inferred magnetic-field orientations at corresponding distances \citep{GonzalezCasanova2019_180510329,2020RNAAS...4..105H}.

The separation of foreground contributions has been particularly valuable in interpreting dust polarization. Toward the LMC, H\,{\sc i}-VChGs were used to estimate the Milky Way foreground magnetic-field orientation, independently validated against starlight polarization, and to facilitate the removal of foreground contamination from Planck 353~GHz dust polarization \citep{Alina2023_220501275}. Toward the supernova remnant W44, H\,{\sc i} and molecular-line VChGs revealed the influence of foreground gas on the observed dust polarization and identified the velocity component most closely associated with the remnant's magnetic-field structure \citep{Liu2022_210913670}. Similarly, toward L1688, combining H\,{\sc i} and $^{12}$CO-VChGs with distance-resolved stellar polarization distinguished magnetic-field contributions from diffuse atomic gas and the molecular cloud, explaining discrepancies between different polarization tracers \citep{Schmaltz2024_231108681}.

Beyond foreground separation, VChGs provide magnetic-field measurements in regions lacking suitable polarization observations, as demonstrated for the Smith Cloud, where H\,{\sc i} VChGs was used to infer both magnetic-field morphology and magnetization \citep{Hu2019_200209948}. Combining velocity-resolved gradients with turbulence-based estimates of magnetic-field strength has further enabled the reconstruction of Galactic magnetic-field orientation and strength as functions of distance \citep{Hu2023_230205047}. More recently, spectroscopic approaches incorporating machine learning have extended these measurements toward the full three-dimensional magnetic-field vector, including its LOS inclination \citep{Hu2025_250507422}. Together, these developments establish velocity-resolved magnetic-field tracing as a means of disentangling LOS structures, mitigating foreground contamination, and reconstructing Galactic magnetic fields in three dimensions.

Synchrotron gradients provide an independent view of the turbulent magnetic field at radio wavelengths. Early polarization-gradient studies showed that spatial fluctuations of the synchrotron Stokes parameters $Q$ and $U$, produced and modified by Faraday rotation in the magnetized ionized medium, encode the properties of MHD turbulence
\citep{Gaensler2011_11102896,Burkhart2012_11113544,Iacobelli2014_14046077, Robitaille2017_170304469,Herron2017_161203992}. SIGs then showed that total synchrotron intensity can recover the POS field orientation \citep{Lazarian2017_170107883}, while SPGs use frequency-dependent synchrotron polarization to exploit Faraday depolarization for magnetic tomography \citep{Lazarian2018_180200028}. Their relation to conventional Faraday tomography and their dependence on magnetization, observing frequency, and Faraday rotation have been explored numerically \citep{Ho2019_190107731,Zhang2019_190411939,Zhang2019_191002378,
Zhang2020_200405270,Wang2021_210603573,Carmo2020_abc331}. These studies also show that synchrotron gradients remain useful in regimes affected by fast-mode intermittency or super-Alfv\'enic turbulence
\citep{Ho2021_210408627,Ho2025_241202102}.

\subsection{Magnetic field in star-forming regions}
\label{sec:applications_starformation}

\begin{figure*}[htbp!]
\includegraphics[width=1.0\linewidth]{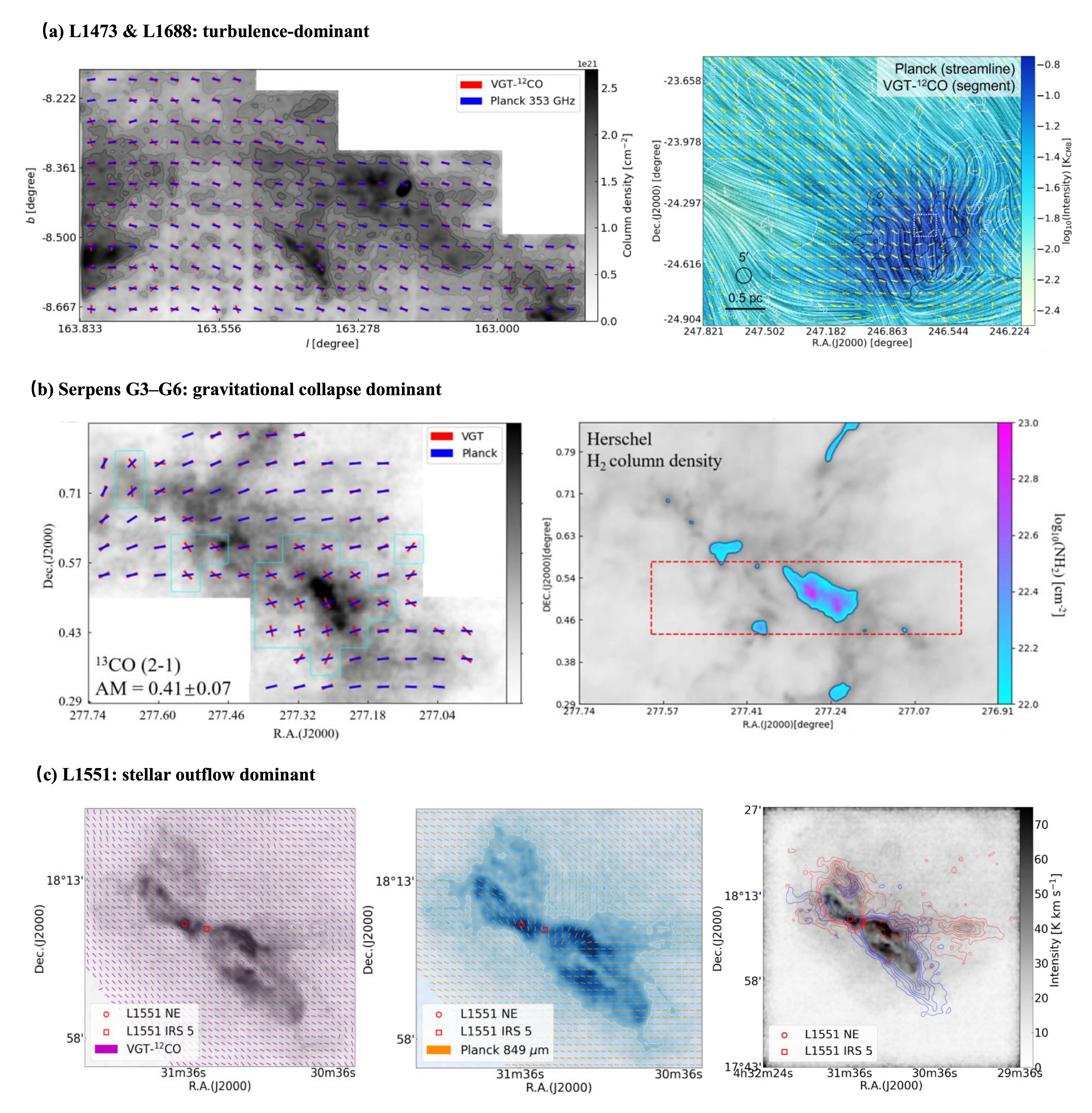}
\caption{
Examples of magnetic-field tracing with velocity gradients in molecular clouds under different dynamical conditions. 
(a) Turbulence-dominated regions L1473 and L1688. Magnetic-field orientations inferred from $^{12}$CO-VGT-VChGs (red segments in the left panel and yellow segments in the right panel) closely follow those derived from Planck 353~GHz ($\simeq850\,\mu{\rm m}$) dust polarization (blue segments or streamlines). Adapted from Schmaltz et al. \cite{2023MNRAS.523.1853S} and \cite{2023MNRAS.524.4431H}. (b) Serpens G3--G6, where self-gravity becomes dynamically important. $^{13}$CO-VGT-VChG orientations (red) show systematic departures from the Planck 353~GHz magnetic field (blue) in the central gravitational-collapse regions (colorful), as determined by the corresponding Herschel H$_2$ column-density map. Adapted from Hu et al. \cite{2021ApJ...912....2H}. (c) L1551, where stellar feedback dominates the local dynamics. The $^{12}$CO-VGT-VChGs magnetic morphology (magenta) is compared with that inferred from Planck 353~GHz ($849\,\mu{\rm m}$) dust polarization (orange). Red- and blueshifted CO emission in the right panel traces the bipolar molecular outflow. Adapted from Liu et al. \cite{2024MNRAS.530.1066L}.}
\label{fig:mc}
\end{figure*}

In relatively diffuse, turbulence-dominated molecular gas, the velocity field remains largely controlled by anisotropic MHD turbulence. Numerical studies show that VCGs preserve the expected alignment over a broad range of sonic and Alfv\'enic Mach numbers, while radiative-transfer calculations demonstrate that useful gradient information survives CO self-absorption and the H\,{\sc i}--H$_2$ transition \citep{GonzalezCasanova2017_160806867,Yuen2017_170107944, Hu2018_180208772,DiegoFGonzalezCasanova2019_170303035,
Hsieh2019_180909806,Hu2020_200800387}. Observational comparisons span a wide range of molecular-cloud environments and polarization tracers. As shown in Figure~\ref{fig:mc}, in nearby Gould Belt clouds, VChG-inferred magnetic fields were compared with Planck
353~GHz ($\simeq850\,\mu{\rm m}$) thermal-dust polarization \citep{Hu2019_200209948}. In Vela C, VCGs and VChGs derived from multiple molecular tracers were compared with BLASTPol 500 $\mu$m dust polarization \citep{Hu2019_190404391}, while thin-channel gradients in Taurus and multi-line VGT measurements in Orion A were compared with Planck 353~GHz polarization \citep{Heyer2020_200610775,Zhao2022_220606717}. These comparisons show that velocity-gradient measurements can recover the POS magnetic-field morphology from diffuse molecular gas to denser cloud structures. Comparing IGs, VCGs, and VChGs additionally helps identify localized shocks and
converging flows, as demonstrated in Monoceros OB1 East \citep{Alina2022_200715344}.

As self-gravity becomes dynamically important, the velocity field departs systematically from the turbulence-dominated state. Simulations predict an approximately $90^\circ$ reorientation of velocity gradients, accompanied by
changes in gradient amplitude and morphology, while IGs respond to the increasing dominance of compressed density structure \citep{KaHoYuen2017_170303026,Hu2020_200206754}. These signatures remain identifiable in self-absorbing molecular lines \citep{Hu2021_200105648}. Observationally, the predicted transition has been identified with VChGs in Serpens G3--G6, L1478/L1482, and L1688 (see Figure~\ref{fig:mc}; \citep{Hu2021_210206225,Schmaltz2023_221012518,Hu2023_221011023}). 

Toward dense cores and cluster-forming regions, turbulence, gravity, and multiple velocity components become increasingly intertwined. In NGC~1333, multi-line VGT measurements using $^{12}$CO, $^{13}$CO, C$^{18}$O, and N$_2$H$^+$ were compared with SOFIA/HAWC+ 214 $\mu$m and JCMT/POL-2 850 $\mu$m dust polarization, probing magnetic-field morphology over different density regimes and spatial scales \citep{Soam2024_241013350}. NH$_3$ VCGs, together with dust-emission and column-density gradients, reveal further scale- and environment-dependent variations in the same region \citep{Chen2024_240718375}. Similar evolution is seen from molecular-cloud envelopes to infrared-dark clumps and, with ALMA, to the G327.29 protocluster, where turbulence and gravitational inflow jointly shape the magnetic morphology \citep{Ramkumar2026_260623802,AKoley2026_260113473}.

Shocks and stellar feedback provide additional dynamical effects. Shock compression can reorganize both the velocity-gradient and magnetic-field geometry \citep{GrandaMunoz2025_240508702}, while protostellar outflows generate coherent high-velocity structures. In L1551, $^{12}$CO-VChGs were compared with magnetic fields inferred independently from Planck 353~GHz ($\simeq849\,\mu{\rm m}$), SOFIA/HAWC+ 214 $\mu$m, and JCMT/SCUPOL 850 $\mu$m dust polarization. The comparison shows a systematic change in gradient--field orientation toward the outflow-dominated central region (see Figure~\ref{fig:mc}; \citep{Liu2024_230904173}). On larger scales, expanding H\,{\sc ii} regions and other compressive flows similarly reshape the gas and field, as seen in G37 and the ring-like G111 cloud
\citep{Zhao2025_250303219,Alina2025_250719281}. Giant molecular filaments show the same coexistence of coherent large-scale fields with localized disruptions associated with collapse or feedback \citep{Zhao2024_231102931}.

\subsection{Magnetic field in the Galactic center}
\label{sec:applications_GC}
\begin{figure*}[htbp!]
\includegraphics[width=1.0\linewidth]{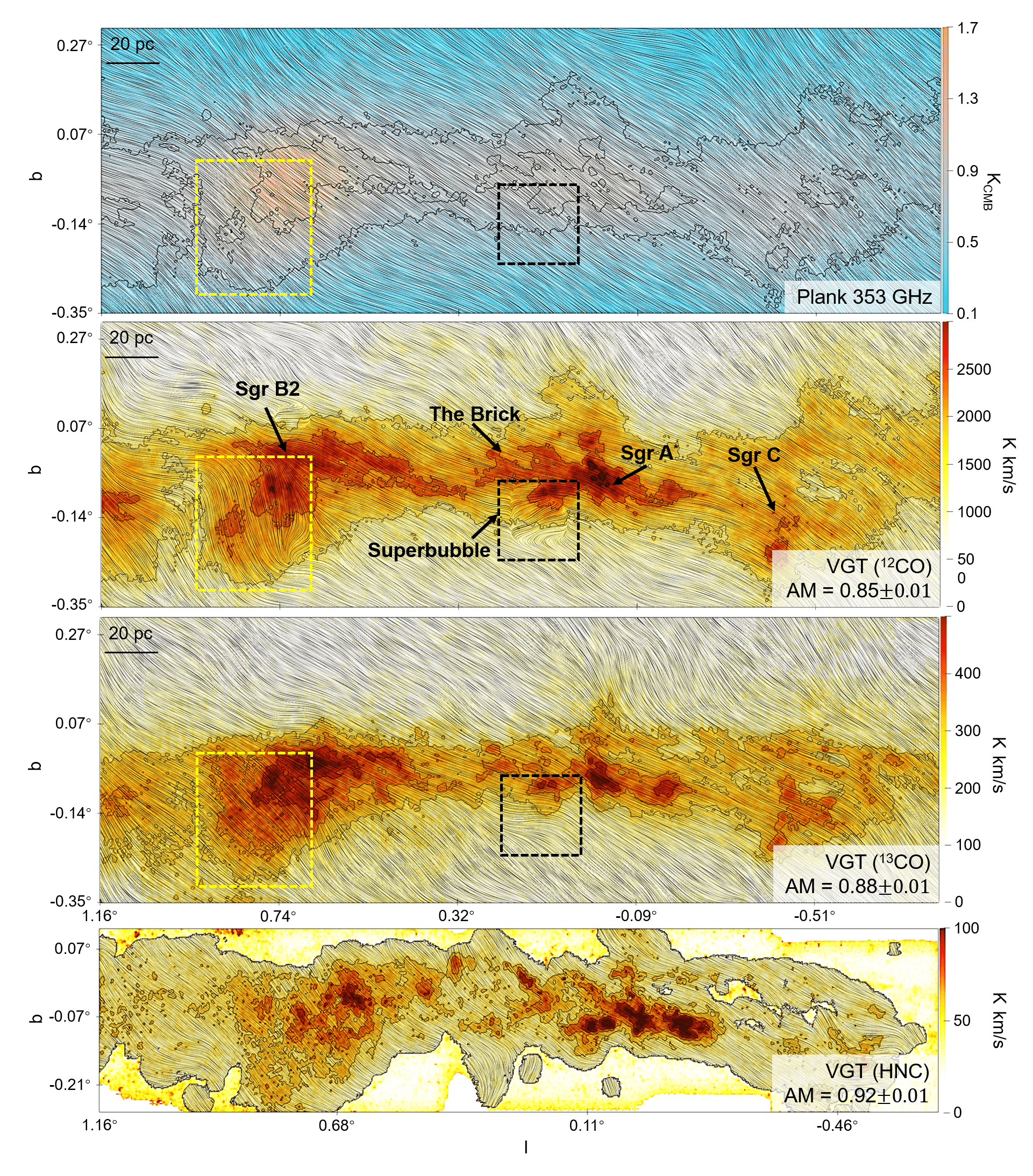}
\caption{Visualization of magnetic fields towards the CMZ using the line integral convolution. The magnetic fields were inferred from Planck 353~GHz polarized dust emission (top) and the VGT-VChGs using $^{12}{\rm CO}$, $ ^ {13} {\rm CO} $, and HNC emissions. We subtracted the foreground contribution from the Planck polarization and overlaid it on the total dust emission intensity map. The VGT-VChGs measurements are overlaid on corresponding integrated intensity colour maps. The contours on $^{12}{\rm CO}$ and $^{13}{\rm CO} $ intensity maps start from twice the mean intensity. The $^{12}{\rm CO}$ contours are also overlaid on the Planck map. Adapted from Hu et al. \citep{Hu2022_210503605}.}
\label{fig:cmz}
\end{figure*}
The Central Molecular Zone (CMZ) combines dense molecular gas, ionized streams, non-thermal plasma, strong shear, and energetic feedback, making multiphase diagnostics essential. VChGs from molecular and ionized lines, IGs from infrared and recombination-line emission, and SIGs from the 1.4~GHz radio continuum recover complementary magnetic structures across scales from $\sim10$ pc to $\sim0.1$ pc \citep{Hu2022_210503605}. 

On large scales, the gradient-inferred magnetic field is broadly consistent with Planck 353~GHz thermal-dust polarization, while toward compact structures such as the Radio Arc and Sgr~A West it agrees with HAWC+ 53 $\mu$m dust polarization (see Figure~\ref{fig:cmz}). In particular, the magnetic field inferred from [Ne\,{\sc ii}]-VChGs toward Sgr~A West closely follows the HAWC+ polarization morphology. Localized departures from the turbulence-dominated alignment reveal regions where coherent flows and other organized dynamics become important.

The severe LOS confusion toward the Galactic center can be reduced with velocity-resolved VChGs. Decomposing $^{12}$CO and $^{13}$CO emission into individual velocity components separates magnetic structures that are superposed in the LOS-integrated Planck 353~GHz polarization map \citep{Hu2022_220107970}. Although the velocity-integrated VChGs field is globally consistent with the Planck polarization, individual components reveal distinct magnetic configurations toward Sgr~B2, the Sgr~A* outflow region, and high-velocity structures associated with the expanding molecular ring. The CMZ therefore demonstrates how multiphase VChGs and velocity decomposition can extend conventional projected polarization measurements toward velocity-resolved magnetic tomography.

\subsection{Magnetic field in external galaxies}
\label{sec:applications_external}
Applying gradient techniques on galactic scales is more challenging than in idealized turbulent systems because the gas dynamics are simultaneously shaped by differential rotation, shear, spiral structure, bars, inflows, stellar feedback, and galaxy interactions. The applicability of the VGT in such complex environments was tested numerically by \citet{Hu2025_241008157} using a cosmological galaxy-merger simulation, which showed that velocity fluctuations remain statistically anisotropic with respect to the local magnetic field throughout the pre-merger, merger, and post-merger stages, providing the physical basis for extragalactic applications. 
\begin{figure*}[htbp!]
\includegraphics[width=1.0\linewidth]{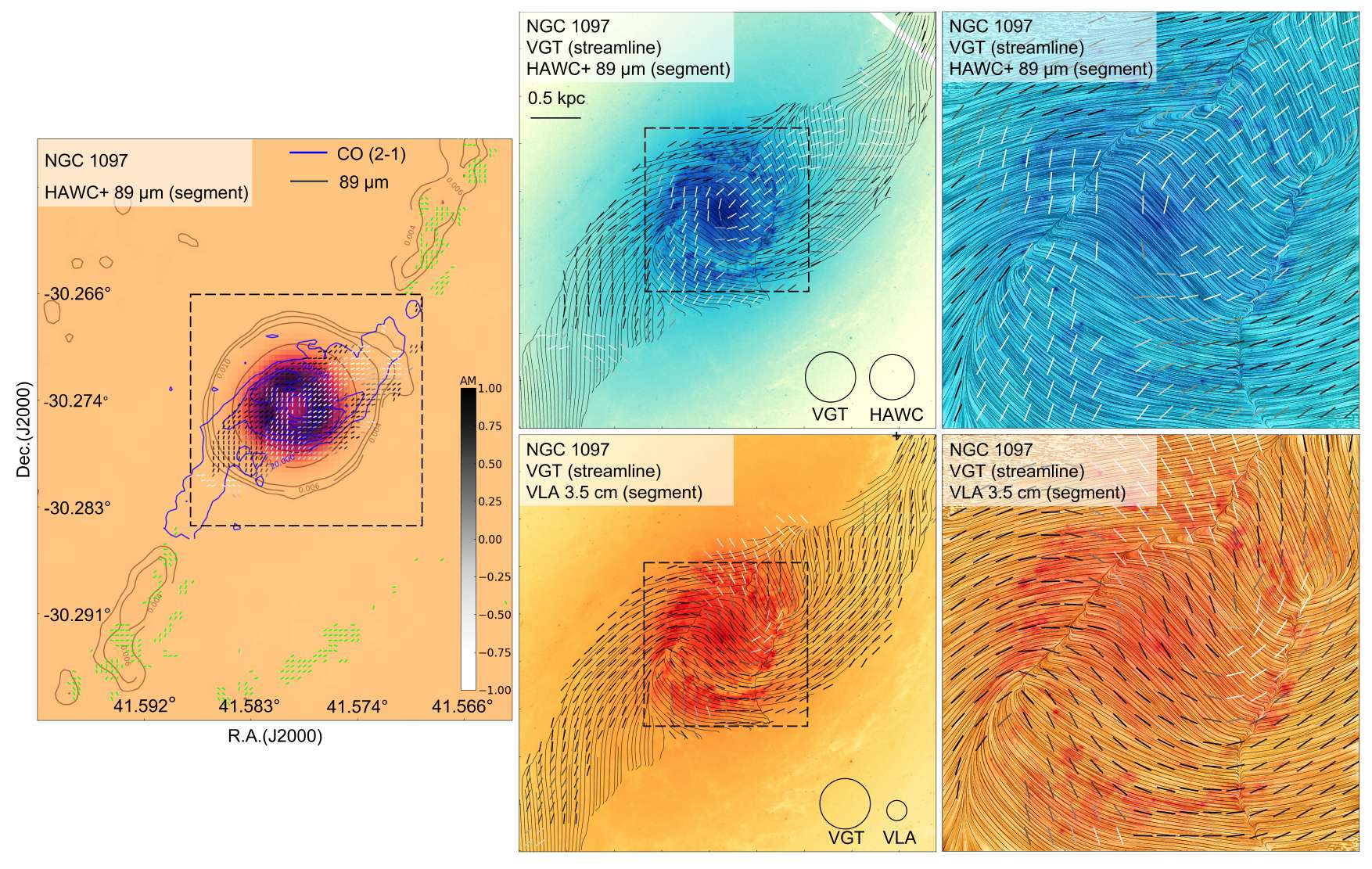}
\caption{Comparison of magnetic-field morphology in NGC~1097 inferred from CO-VChGs and independent polarization measurements. The left panel shows the large-scale field traced by $^{12}$CO$(2$--$1)$ VGT-VChGs together with SOFIA/HAWC+ 89~$\mu$m dust polarization; the color of the polarization segments indicates the local alignment measure between the two tracers. The upper middle and right panels zoom into the circumnuclear region and starburst ring, comparing the VGT-VChGs field (black streamlines) with HAWC+ 89~$\mu$m polarization (white segments). The lower panels show the corresponding comparison with VLA+Effelsberg 3.5~cm synchrotron polarization. The broad agreement across molecular-gas, far-infrared dust, and radio-synchrotron tracers demonstrates that velocity gradients recover the large-scale magnetic morphology of the galaxy, while localized departures in the nuclear ring and inner bar trace regions where organized galactic dynamics become important. Adapted from Hu et al. \citep{Hu2022_220605423}.}
\label{fig:NGC1097}
\end{figure*}

Observationally, CO-VChGs have been compared with independent magnetic-field measurements from both far-infrared dust polarization and radio synchrotron polarization. In nearby Seyfert galaxies, the VGT-inferred molecular-gas fields broadly follow the polarization-inferred field while revealing systematic changes in spiral arms, bars, and nuclear regions \citep{Hu2022_220605423}. Figure~\ref{fig:NGC1097} illustrates this comparison for NGC~1097, where the $^{12}$CO-VChGs show broad agreement with both SOFIA/HAWC+ 89~$\mu$m dust polarization and VLA+Effelsberg 3.5~cm synchrotron polarization, while localized departures around the circumnuclear ring and inner bar reveal substantial non-azimuthal magnetic components associated with bar-driven inflow and nuclear dynamics. 

Similar comparisons extend to multiphase and feedback-regulated systems and provide information beyond magnetic-field morphology alone. In NGC~3627, CO and H$\alpha$ VChGs probe the magnetic fields associated with molecular and warm-ionized gas, respectively, while VLA+Effelsberg 8.46~GHz synchrotron polarization traces the more diffuse magnetized ISM and HAWC+ 154~$\mu$m polarization traces dust-bearing gas in star-forming regions \citep{Liu2022_220806090}. Differences among these tracers reveal how the magnetic field couples differently to individual gas phases and how tidal interaction and large-scale gas flows distort the field in different components of the ISM. 

In 30~Doradus, [C\,{\sc ii}] and CO-VChGs probe the atomic/molecular interface and denser molecular gas, while HAWC+ 89, 154, and 214~$\mu$m polarization provides an independent view of the dust-traced field across the star-forming complex \citep{Tram2023_220512084}. Their relative orientations distinguish regions where the field follows the turbulent gas from those where expanding shells, gravitational contraction, or stellar feedback reorganize both the gas and magnetic geometry. 

In NGC~628, CO-VChGs trace the field associated with the molecular disk, whereas JVLA 3.1~GHz synchrotron polarization samples the large-scale field in the diffuse relativistic plasma \citep{Zhao2024_240402651}. The agreement between the two indicates magnetic coherence across distinct ISM phases, while the systematic curvature of the field around large bubbles suggests compression and deflection by feedback-driven expansion.

In Centaurus~A, CO-VChGs reproduce the magnetic morphology traced by HAWC+ 89~$\mu$m dust polarization across much of the warped molecular disk, implying that the molecular gas and dust remain coupled to a common large-scale field \citep{Nguyen2024_240604242}. Localized departures near the jet and outflow instead identify regions where jet--ISM interaction and coherent gas motions overwhelm the standard turbulent anisotropy. These observations therefore show that multi-tracer gradient measurements can diagnose not only magnetic-field geometry, but also the coupling of magnetic fields across ISM phases and their reorganization by interactions, compression, feedback, and outflows.

\subsection{Magnetic field in galaxy clusters}
\label{sec:applications_clusters}
Galaxy clusters provide a distinct regime in which weakly compressible, frequently super-Alfv\'enic turbulence coexists with strong LOS averaging and Faraday depolarization. Early applications of gradient techniques to the ICM used Chandra X-ray intensity fluctuations and radio synchrotron emission to probe magnetic morphology in both cool-core and non-cool-core systems \citep{Hu2020_200706219}. In the cool-core clusters Perseus, M87, and Abell~2597, the inferred magnetic field shows a preferentially tangential orientation with respect to the cluster center, whereas the non-cool-core Coma cluster exhibits a substantially weaker preferred orientation. The field also follows sloshing structures and shows localized draping around AGN-inflated bubbles and merging substructures, indicating that the magnetic topology reflects the combined influence of turbulence, stratification, and coherent gas motions. The preferentially tangential configuration in cool cores has subsequently been confirmed statistically in cosmological MHD simulations, which show a characteristic tangential magnetic bias in cool-core clusters near $\sim0.1\,r_{500}$ that is largely absent in non-cool-core systems \citep{Lehle2025_250712517}. This establishes the tangential field as a systematic property of cool-core cluster dynamics rather than a feature of individual systems.

\begin{figure*}[htbp!]
\includegraphics[width=1.0\linewidth]{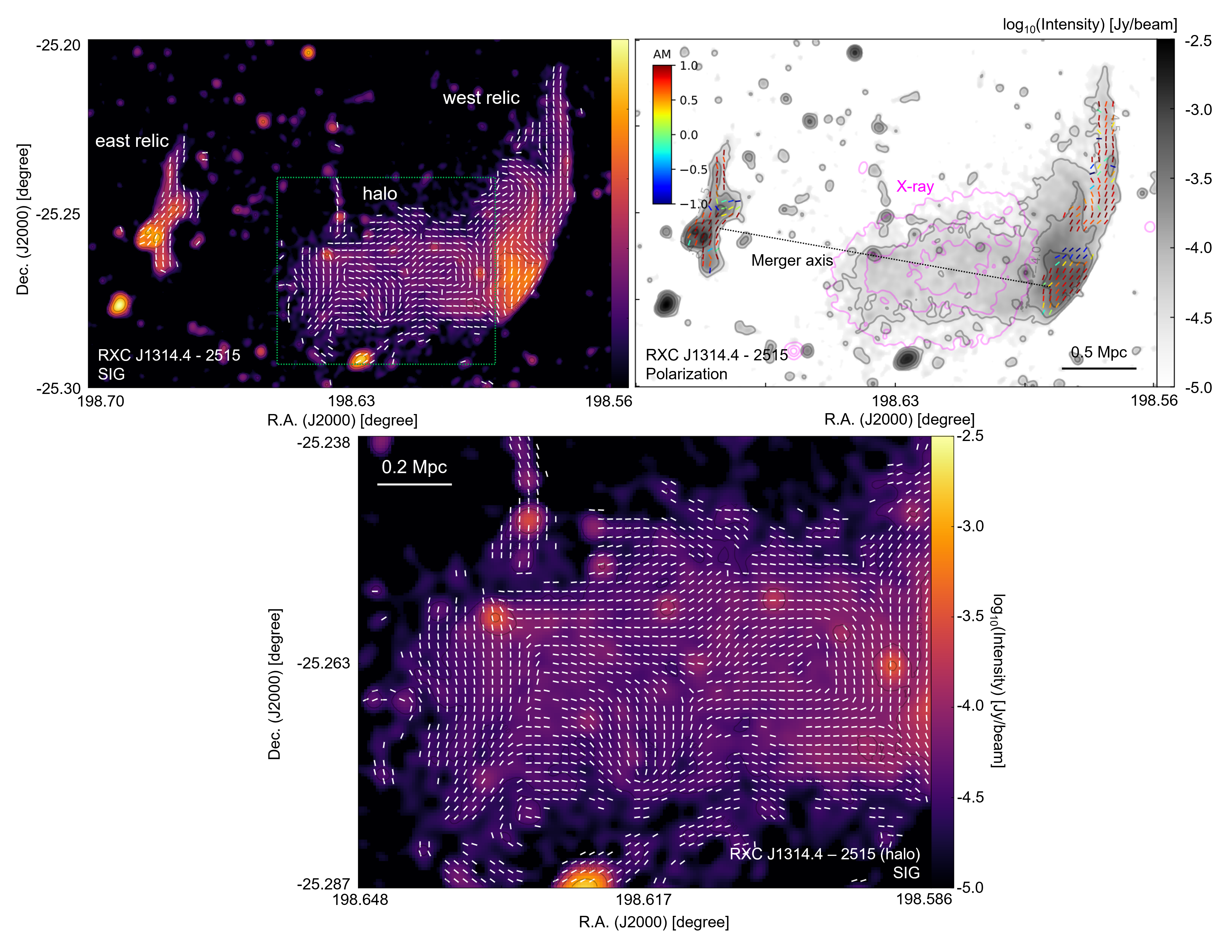}
\caption{Magnetic-field morphology in the merging galaxy cluster RXC~J1314.4$-$2515 inferred from SIGs and synchrotron polarization. Top left: POS magnetic-field orientations derived with SIG (white segments) across the eastern and western radio relics and the central radio halo, overlaid on the MeerKAT 1.28~GHz synchrotron-intensity map. The effective SIG resolution is approximately $25''$, corresponding to $\sim120$~kpc. Top right: magnetic-field orientations inferred independently from JVLA 3~GHz synchrotron polarization in the radio relics. The colored polarization segments indicate the local alignment measure between the polarization- and SIG-inferred fields; magenta contours show the XMM--Newton X-ray emission and the dotted line marks the merger axis. Bottom: enlarged view of the central radio halo, where Faraday depolarization strongly suppresses detectable polarization, but SIG continues to recover the magnetic-field morphology. Adapted from Hu et al. \citep{Hu2024_230610011}.}
\label{fig:cluster_SIG}
\end{figure*}

Radio synchrotron gradients provide a complementary probe of cluster magnetic fields. SIGs are particularly advantageous because they infer magnetic-field anisotropy from total synchrotron intensity and therefore do not require detectable polarized emission. Numerical studies show that magnetic and velocity fluctuations retain measurable anisotropy over a broad range of ICM magnetization and that SIGs remain statistically correlated with the magnetic field even in super-Alfv\'enic turbulence \citep{Hu2024_230610011,Ho2025_241202102,Hu2025_250508275}. Observationally, SIGs applied to JVLA and MeerKAT radio halos and relics have mapped coherent magnetic structures in five dynamically disturbed clusters \citep{Hu2024_230610011}. Figure~\ref{fig:cluster_SIG} illustrates the case of RXC~J1314.4$-$2515. In the peripheral radio relics, where polarized synchrotron emission remains detectable, the SIG-inferred magnetic-field orientations agree closely with those measured independently from JVLA 3~GHz synchrotron polarization, providing a direct observational validation of the method. In the central radio halo, by contrast, strong Faraday depolarization suppresses the polarized signal while the total synchrotron intensity remains observable; SIGs therefore recover magnetic-field structure that is largely inaccessible to conventional polarimetry. The halo field displays large-scale organization related to the merger geometry together with smaller-scale turbulent structure, directly connecting magnetic morphology to merger-driven gas motions.

Ultra-deep LOFAR observations of Abell~2255 extend this approach from the central radio halo through bridges, filaments, and peripheral relics to approximately the virial radius \citep{Botteon2026_260714209}. These synchrotron structures probe different dynamical environments: the central halo traces volume-filling merger-driven turbulence, relics reveal magnetic compression and ordering associated with shocks, while bridges and diffuse peripheral emission probe large-scale flows and magnetized plasma in the cluster outskirts. Gradient techniques therefore provide not only projected maps of cluster magnetic fields but also a means of connecting magnetic topology to cool-core stratification, turbulence, shocks, mergers, AGN feedback, and large-scale gas flows across the ICM.

\section{Conclusions and outlook}
\label{sec:conclusion}

Gradients of turbulent fluctuations provide a unified framework for probing astrophysical magnetic fields. Their physical basis lies in the scale-dependent anisotropy of MHD turbulence: Alfv\'enic and slow-mode fluctuations cascade preferentially perpendicular to the local magnetic field rather than along it, producing gradients that are preferentially perpendicular to the field. This behavior applies to velocity and magnetic-field fluctuations and, under appropriate conditions, to density fluctuations as well. VCGs, VChGs, IGs, SIGs, and SPGs probe different observables, but all exploit this common connection between turbulent anisotropy and magnetic geometry.

Intermittent fast modes, shocks, self-gravity, outflows, shear, and other dynamical motions can compete with turbulent fluctuations, while thermal broadening, opacity, and self-absorption modify how those fluctuations are mapped into observables. Numerical studies show that these effects may produce changes in gradient orientation and amplitude. Departures from the standard alignment can therefore carry dynamical information rather than simply representing a limitation of the technique, providing diagnostics of gravitational collapse, shock compression, coherent flows, and stellar feedback.

Observational applications now span the diffuse ISM, star-forming regions, the Galactic center, external galaxies, and galaxy clusters. Spectroscopic VChGs provide velocity-resolved magnetic information and can separate structures overlapping along the LOS, while SIGs and SPGs probe magnetic-field fluctuations and, through Faraday effects, different effective depths within magnetized plasmas. Together with polarization, distance information, and turbulence-based estimates of magnetization, these measurements are extending magnetic-field studies from projected morphology toward three-dimensional tomography.

A major opportunity is to use gradient observables as dynamical as well as magnetic diagnostics. In star-forming regions, the relative behavior of velocity, density, and magnetic-field gradients can distinguish turbulence, gravitational collapse, shocks, and feedback, offering a way to follow the dynamical evolution from cloud to core scales. In the diffuse Galactic ISM, three-dimensional magnetic-field reconstructions can improve models of polarized dust and synchrotron foregrounds and their LOS decorrelation, with direct relevance to precision CMB polarization measurements. Magnetic-field and magnetization maps can also provide physically motivated constraints on charged-particle transport, including Galactic cosmic rays and the deflection of ultra-high-energy cosmic rays. At larger scales, future low-frequency radio and spectroscopic surveys may extend gradient techniques to cluster outskirts, intergalactic filaments, and the magnetized cosmic web, where conventional polarimetry becomes increasingly difficult.

Realizing these applications will require more realistic modeling of multiphase turbulence, radiative transfer, self-gravity, shocks, feedback, and cosmic-ray transport, together with improved statistical approaches such as sub-block averaging, velocity decomposition, scale filtering, and machine-learning-assisted analysis. By exploiting the statistical anisotropy of turbulent fluctuations while simultaneously using departures from the standard alignment as signatures of non-turbulent dynamics, gradient techniques offer a framework for connecting magnetic structure to the physical processes shaping the magnetized Universe.

\vspace{6pt} 





\authorcontributions{Conceptualization, Y.H.; writing---original draft preparation, Y.H.; writing---review and editing, Y.H.; visualization, Y.H. The author has read and agreed to the published version of the manuscript.}

\funding{The Sherman Fairchild Postdoctoral Fellowship at the California Institute of Technology supports Y.H.}

\dataavailability{No new data were created or analyzed in this study. Data sharing is not applicable to this article.}

\acknowledgments{The author acknowledges the use of SDSC Expanse and NCSA Delta through ACCESS allocations PHY230032, PHY230033, PHY230091, PHY230105, PHY230178, PHY240188, and PHY240183. ACCESS is supported by National Science Foundation grants \#2138259, \#2138286, \#2138307, \#2137603, and \#2138296.}

\conflictsofinterest{The author declares no conflicts of interest.} 



\abbreviations{Abbreviations}{
The following abbreviations are used in this manuscript:
\\
\noindent
\begin{tabular}{@{}ll}
MHD & Magnetohydrodynamic\\
ISM & Interstellar Medium\\
ICM & Intracluster Medium\\
POS & Plane of Sky\\
LOS & Line of Sight\\
PPV & Position--Position--Velocity\\
VGT & Velocity Gradient Technique\\
VCG & Velocity Centroid Gradient\\
VChG & Velocity Channel Gradient\\
IG & Intensity Gradient\\
SIG & Synchrotron Intensity Gradient\\
SPG & Synchrotron Polarization Gradient\\
VDA & Velocity Decomposition Algorithm\\
CMZ & Central Molecular Zone\\
CNM & Cold Neutral Medium
\end{tabular}
}




\begin{adjustwidth}{-\extralength}{0cm}

\reftitle{References}


\bibliography{sample631}

\PublishersNote{}
\end{adjustwidth}
\end{document}